\documentclass[journal]{IEEEtran}

\usepackage{todonotes}
\setuptodonotes{fancyline,color=blue!30}
\usepackage{hyperref}
\usepackage{amsmath}
\usepackage{graphicx}
\usepackage[ruled,vlined,linesnumbered]{algorithm2e}
\usepackage{times} 
\usepackage{bbm}
\usepackage{amssymb}  
\usepackage{xcolor}
\usepackage{float}
\usepackage{mathtools}
\usepackage{amsthm}
\usepackage{cleveref}
\usepackage{bbold}
\usepackage{enumitem}
\usepackage[detect-weight=true, detect-family=true]{siunitx}
\usepackage{cite}

\usepackage[english]{babel}
\usepackage[autostyle]{csquotes}

\newtheorem{theorem}{Theorem}

\newtheorem{remark}{Remark}
\newtheorem{proposition}{Proposition}

\newcommand{\argmax}{\text{argmax}}

\newcommand{\vQ}{\mathbf{Q}}

\newcommand{\vR}{\mathbf{R}}
\newcommand{\vr}{\mathbf{r}}

\newcommand{\vv}{\mathbf{v}}
\newcommand{\vp}{\mathbf{p}}
\newcommand{\vw}{\mathbf{w}}
\newcommand{\ve}{\mathbf{e}}
\newcommand{\vd}{\mathbf{d}}
\newcommand{\vu}{\mathbf{u}}

\newcommand{\vx}{\mathbf{x}}
\newcommand{\vc}{\mathbf{c}}
\renewcommand{\vd}{\mathbf{d}} 

\newcommand{\vz}{\mathbf{z}}

\newcommand{\oT}{\mathcal{T}}
\newcommand{\trace}{\ \mathrm{trace} }

\newcommand{\mA}{A}
\newcommand{\mK}{K}
\newcommand{\mH}{H}
\newcommand{\mX}{X}
\newcommand{\mF}{F}
\newcommand{\mB}{B}
\newcommand{\mM}{M}
\newcommand{\mW}{W}

\newcommand{\fP}{\textit{P}}
\newcommand{\fR}{\textit{R}}
\newcommand{\fQ}{\textit{Q}}

\newcommand{\ff}{\textit{f}}

\newcommand{\feta}{\eta}

\newcommand{\fr}{\textit{r}}
\newcommand{\ffpi}{\Omega^*}
\newcommand{\fJ}{\textit{J}}
\newcommand{\fPhi}{\Phi}

\newcommand{\sS}{\mathcal{S}}
\newcommand{\sA}{\mathcal{A}}
\newcommand{\sI}{\mathcal{I}}

\newcommand{\sOmega}{\Omega}
\newcommand{\sM}{\mathcal{M}}

\newcommand{\td}{^{\mathrm{c}}}
\newcommand{\bellman}{^{\mathrm{b}}}

\newcommand{\eR}{\mathbbm{R}} 
\newcommand{\Z}{\mathbbm{Z}} 
\newcommand{\ds}{\mathrm{ds}}

\renewcommand{\th}{^{\text{th}}}

\newcommand{\tempA}{\mathbbm{A}}

\newcommand{\pP}{\mathbbm{P}} 

\newcommand{\I}{\mathbbm{I}}

\begin{document}

\title{H-TD\textsuperscript{2}: Hybrid Temporal Difference Learning for Adaptive Urban Taxi Dispatch} 
\author{Benjamin~Rivière and~Soon-Jo~Chung}

\maketitle

\begin{abstract}
We present H-TD\textsuperscript{2}: \underline{H}ybrid \underline{T}emporal \underline{D}ifference Learning for \underline{T}axi \underline{D}ispatch, a model-free, adaptive decision-making algorithm to coordinate a large fleet of automated taxis in a dynamic urban environment to minimize expected customer waiting times. 
Our scalable algorithm exploits the natural transportation network company topology by switching between two behaviors: distributed temporal-difference learning computed locally at each taxi and infrequent centralized Bellman updates computed at the dispatch center. 
We derive a regret bound and design the trigger condition between the two behaviors to explicitly control the trade-off between computational complexity and the individual taxi policy's bounded sub-optimality; this advances the state of the art by enabling distributed operation with bounded-suboptimality. 
Additionally, unlike recent reinforcement learning dispatch methods, this policy estimation is adaptive and robust to out-of-training domain events.
This result is enabled by a two-step modelling approach: the policy is learned on an agent-agnostic, cell-based Markov Decision Process and individual taxis are coordinated using the learned policy in a distributed game-theoretic task assignment.
We validate our algorithm against a receding horizon control baseline in a Gridworld environment with a simulated customer dataset, where the proposed solution decreases average customer waiting time by \SI{50}{\percent} over a wide range of parameters. We also validate in a Chicago city environment with real customer requests from the Chicago taxi public dataset where the proposed solution decreases average customer waiting time by \SI{26}{\percent} over irregular customer distributions during a 2016 Major League Baseball World Series game.
\end{abstract}


\section{Introduction}
\begin{figure}
\begin{center}
    \includegraphics[width=0.85\linewidth,trim={0cm 2.25cm 0cm 3cm},clip]{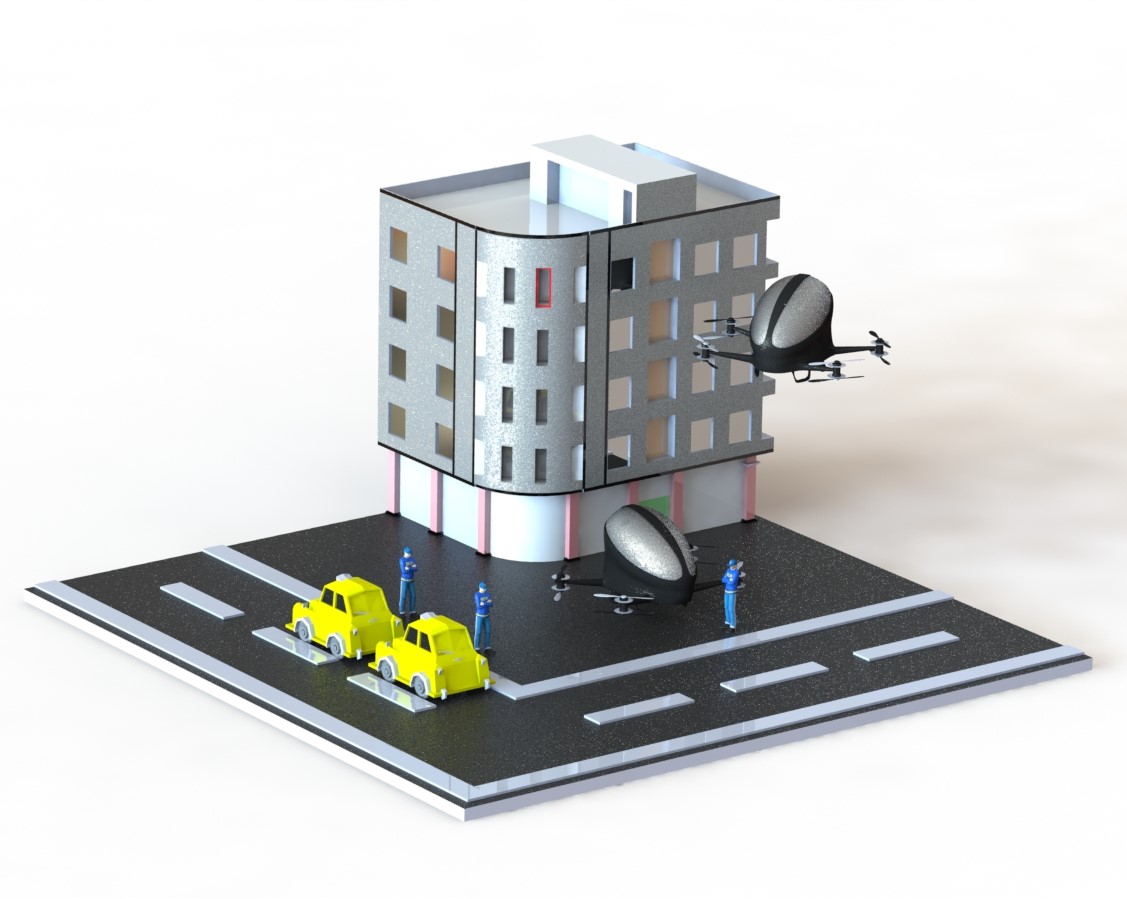}
    \caption{Concept graphic of an intelligent transportation network. Autonomous taxis, that can include both ground and air vehicles, estimate in real-time the customer demand and coordinate locally to behave with bounded sub-optimality. 
    } 
    \label{fig:concept_graphic}
\end{center}
\end{figure}

Coordinating a large fleet of automated taxis in complex and dynamic urban environments is an anticipated challenge for transportation network companies such as Uber, Lyft, Waymo, and Tesla. A typical urban mobility problem for these companies is taxi dispatch, where a fleet of taxis service customers and the remaining, idle taxis are coordinated with a dispatch algorithm to minimize the customer waiting time of future requests. In practice, a transportation network company might be composed of a dispatch center equipped with complete information and a large computational budget and a fleet of taxis, each operating with local information and a limited amount of processing power and communication bandwidth (see Fig.~\ref{fig:concept_graphic}). In this manner, the transportation network company network can be decomposed into an underlying \enquote{star-topology} network between taxis and the dispatch center, and an arbitrary peer-to-peer network between taxis. The proposed algorithm, H-TD\textsuperscript{2}, exploits this topology explicitly by proposing a hybrid algorithm with two distinct behaviors: the central node computes exact, large-batch policies infrequently, and each taxi computes approximate, online updates with local information. 



\begin{figure}
\begin{center}
\includegraphics[width=\linewidth]{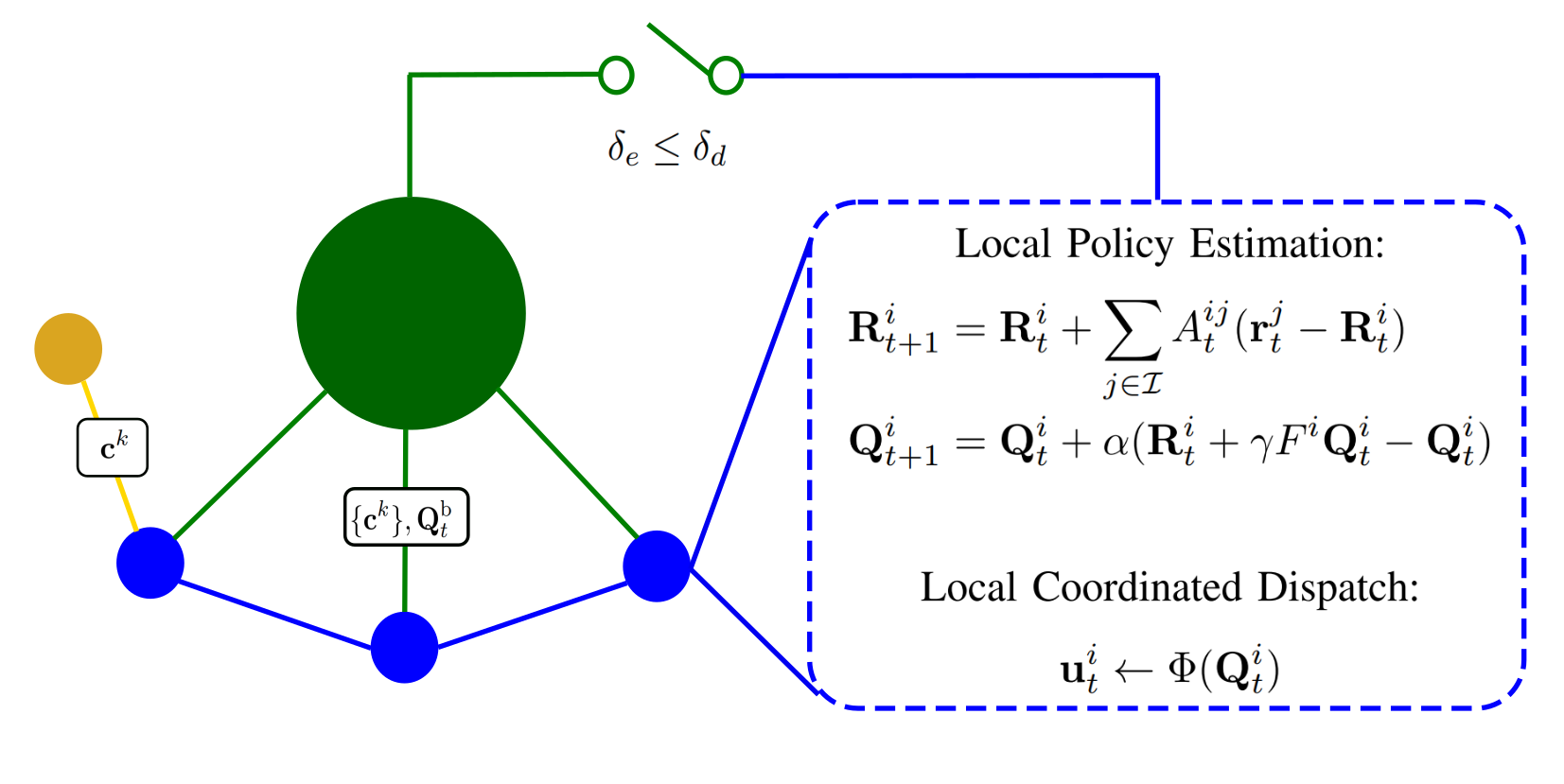}
\caption{Overview of H-TD\textsuperscript{2}, where blue represents the taxi network, yellow represents the customers, and green represents the dispatch center. The i$\th$ taxi estimates the dispatch policy with local operations: distributed estimation of reward, $\vR^i_t$ computed in~\eqref{eq:dtd_reward_estimator}, and temporal difference learning to update the policy, $\vQ^i_t$,~\eqref{eq:dtd}. If any of the taxis determines that its policy estimate error, $\delta_e$, is larger than the user specified threshold, $\delta_d$, the taxi signals to the dispatch center to receive a centralized policy update, $\vQ\bellman_t$~\eqref{eq:bellman}. Finally, each free taxi uses the policy in a game theoretic formulation, $\Phi$~\eqref{eq:marginal_utility}, to find its dispatch position vector, $\vu^i_t$. }
\label{fig:overview}
\end{center}
\end{figure}

The overview of H-TD\textsuperscript{2} is shown in Fig.~\ref{fig:overview}. At a given timestep, the closest taxis service the new customer requests, and the rest of the free taxis are dispatched to reduce expected waiting time of future requests. The free taxis coordinate with a distributed game theoretic scheme to optimize their policy estimate, where the policy is estimated as follows: the servicing taxis communicate the customer data to their neighboring taxis, where the expected reward (e.g. customer waiting time) is estimated with a distributed estimation algorithm. Then, all taxis use the estimated reward to update their policy estimate with temporal difference learning. If any of the taxis determine that their policy estimate error is larger than the user specified threshold, the taxi signals to the dispatch center for a centralized policy update.


The contributions of the paper are stated as follows:
\begin{itemize}[leftmargin=*]
    \item 
    We derive a novel regret bound by leveraging distributed estimation methods in local online policy estimation and introduce a trigger condition to the batch update, permitting the user to explicitly specify the policy's computational and communication expense vs. bounded sub-optimality trade-off. This advances the state of the art by enabling distributed operation with bounded sub-optimality. 
    \item We propose a taxi-dispatch solution that is adaptive, model-free, and coordinated. Unlike state-of-the-art reinforcement-learning dispatch methods, our method directly adapts the policy based on real-time data, thereby providing a property of robustness to irregular urban mobility events such as traffic, weather, and major public events. This advancement is enabled by two step approach: first we propose a hybrid policy estimation in a finite-dimensional, agent-agnostic cell abstraction, and then we interface the resulting policy estimation for agent-based coordination with a local prescriptive game-theoretic task assignment. 
\end{itemize}
We demonstrate the performance and computational properties of our method with numerical experiments: our algorithm reduces customer waiting time compared to a receding horizon control baseline and the simulation runtime is linear with the number of agents. We also validate our claim that adaptive algorithms are robust to general irregular events with a case study of the Chicago City taxis during the 2016 Major League Baseball World Series.

The remainder of the paper is organized as follows: in Sec.~\ref{sec:related_work}, we review the related literature and compare our method with the state of the art. In Sec.~\ref{sec:problem_description}, we present the taxi dispatch problem description and a motivating example. In Sec.~\ref{sec:dispatch_mdp}, we present the cell-based Markov Decision Process (MDP). In Sec.~\ref{sec:algorithm}, we discuss the exact and approximate solutions to the MDP and the integration of the learned policy into a game theoretic method. In Sec.~\ref{sec:numerical_experiments}, we present numerical experiments demonstrating the advantages of our algorithm compared to a receding horizon control baseline in simulated and real customer datasets. The details of the fleet simulator implementation are presented in Appendix~\ref{appendix:fleet_simulator}.

\section{Related Work}
\label{sec:related_work}

Recent urban mobility research has developed dynamic and scalable methods. 
A well-studied example is the vehicle routing and dial-a-ride problems~\cite{Ho_2018,pillac_2013} where taxis find a minimum cost path through a routing graph, and its dynamic extension, where part or all of the customer information is unknown and revealed dynamically. 
Recent dynamic routing research proposes scalable solutions to dynamic routing problems with bio-inspired methods~\cite{claes_2011}, data-driven methods~\cite{menda_2019,abbasi_2019,ke_2019}, and model-based methods~\cite{luo_2018,schutter_2017}. In this paper, we study a variant of the dynamic routing problem, taxi dispatch, where we propose a novel two-stage approach: distributed estimation with temporal difference learning, and game-theoretic coordination. This advances the state of the art by permitting adaptive distributed operation with bounded sub-optimality with respect to the optimal centralized policy. 

Taxi dispatch is an emerging urban mobility problem where free taxis are dispatched to locations in the map to minimize customer waiting time of future requests. Recent approaches have adopted model-based~\cite{miao_2016,zhang_2016} and model-free~\cite{oda_2018} methods. An online model-based method like~\cite{miao_2016} uses real-time data to fit a system prediction model (for example, customer demand and taxi supply) and then compute a receding horizon control solution in response to that model. Pre-specified system models can be over-restrictive, and recent reinforcement-learning model-free methods~\cite{oda_2018,Tang_2019,Xu_2018} have been used successfully to overcome this limitation. In a model-free method, events are not explicitly modelled, they are captured by the arbitrary dynamics of the underlying reward. In regular operation, this reward is periodic, and can be accurately predicted (either explicitly or implicitly) and used for fleet control. However, it is possible that an irregular event occurs out of training domain and causes the reward dynamics to be unpredictable. In this case, we argue that it is better to adapt in real-time than predict with irrelevant data. Our model-free approach adapts the policy directly in response to real-time data, achieving performance that is robust to unpredictable, irregular events such as weather, accidents, and major public events. 

Our method leverages results from reinforcement learning in convergence of temporal difference iteration~\cite{littman_1998,csaji_2008,szepesvari_1999} in a dynamic environment, i.e. when the reward or transition probabilities are changing over time. 
An alternative online model-free approach is online actor-critic~\cite{Vamvoudakis_2010}, where our work differs from this result in two ways: we consider a general non-quadratic reward function and we consider a multi-agent setting by analyzing a hierarchical system of a temporal difference iteration with distributed estimation of the reward model. To the best of the authors knowledge, the only other work to propose a distributed temporal difference algorithm is recent work~\cite{doan_2019} that addresses the convergence properties of consensus on model parameters in the case of linear function approximations. 

In general, multi-agent reinforcement learning research is challenging because the MDP's state and action space dimensionality is coupled to the number of agents, which is typically handled by using either (i) function approximation methods such as deep neural networks or (ii) decoupled, decentralized solutions. A survey paper on multi-agent reinforcement learning discusses additional methods~\cite{schutter_2015}. In contrast to an agent-based (or Lagrangian) approach, our method uses a naturally scalable cell-based (or Eulerian) model that decouples the problem dimensionality from the number of agents, inspired by a method used in probabilistic swarm guidance~\cite{bandyopadhyay_2017}.

Because of the cell-based abstraction, our algorithm requires an additional task assignment component to coordinate taxis. 
Task assignment is a canonical operations research problem and there exists many available centralized~\cite{kuhn_2010, bertsekas_1988, bertsekas_1991,bellingham_2003} and decentralized~\cite{morgan_2016,dionne_2007,sujit_2007}.
Among these options, we use a distributed prescriptive game theory~\cite{marden_2012} approach that leverages existing asymptotic game theoretic optimality and convergence results. In contrast to conventional \emph{descriptive} game theory, \emph{prescriptive} game theory designs multi-agent local interactions to achieve desirable global behavior. Using one such method, binary log-linear learning~\cite{marden_2012}, the taxis achieve global cooperative behavior with only local information. 

\section{Problem Description}
\label{sec:problem_description}

\textit{Notation}: We denote vectors with a bold symbol, matrices with plain uppercase, scalars parameters with plain lowercase, functions with italics, and we use caligraphic symbols for operators and sets. We denote a taxi index with an $i$ or $j$ superscript, a customer index with a $k$ superscript, and the time index with a subscript $t$. Also, $I_n$ denotes the $n$-dimensional identity matrix.

\textit{Problem Statement}:
We consider the urban taxi dispatch problem, where we control a fleet of taxis to minimize customer waiting time. At each timestep, each customer requests is serviced by the nearest taxi. These \emph{servicing} taxis use customer information to update their reward model and exchange information with neighboring taxis. The remaining \emph{free} taxis are dispatched to locations in the map according to the proposed dispatch algorithm. 
The overall fleet control is summarized in Algorithm~\ref{algo:fleet_control} and implementation details are given in Appendix~\ref{appendix:fleet_simulator}. 

\begin{algorithm}
    \label{algo:fleet_control}
    \SetAlgoLined
    initialize taxi fleet\;
    \For{$t \in [t_0:t_f]$}{
        broadcast local customer requests to taxis\;
        assign closest free taxis to service customers\;
        dispatch free taxis to locations in the map\;
    }
\caption{Fleet Control Problem}
\end{algorithm}

\begin{figure}
    \centering
    \includegraphics[width=0.85\linewidth,]{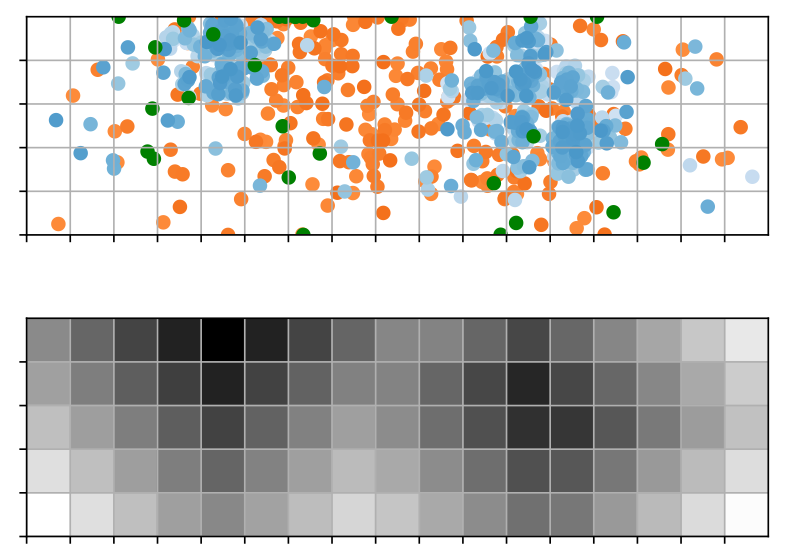}    
    \caption{State space representation of a Gridworld simulation with 1000 taxis, with corresponding value function estimation. In the top subplot, the blue dots are free taxis positions, the orange dots are positions of taxis currently servicing customers, and the green dots are the new customers requests pickup positions. In the bottom subplot, the approximate value function distribution is shown over the state space. }
    \label{fig:gridworld_smallscale_statespace}
\end{figure}

The system, as shown in Fig.~\ref{fig:gridworld_smallscale_statespace}, is composed of customers and taxis. 
The $k\th$-customer state, $\vc^k$, is composed of the time of request, trip duration, pickup location, and dropoff location, i.e. $\vc^k = [t^k_r,t^k_d,\vp^{k,p},\vp^{k,d}]$ and its pickup location is shown in green in the top subplot. 
The $i\th$-taxi is defined by a position vector, $\vp^i_t$ and an operation mode: free (shown in blue) or servicing (shown in orange). 
The dispatch solution is a desired position vector for each of the free taxis, $\vu^i_t$, that results in minimizing customer waiting time over a time horizon. 
Our method estimates the optimal policy, visualized with the value function over the state-space in the bottom subplot, that maximizes the expected reward over time.

\section{Cell-Based Markov Decision Process}
\label{sec:dispatch_mdp}
The previous section described the dispatch problem with a agent-based perspective, i.e. in terms of positions and actions of individual taxis and customers. Next, we will introduce the cell-based Markov Decision Process (MDP), where cell-based refers to an Eulerian perspective in which we analyze values like location and reward with respect to cells of a discretized map as shown in Fig.~\ref{fig:gridworld_smallscale_statespace}. The cell-based formulation decouples the decision making problem from the number of agents, permitting a finite dimensional policy representation. The decision making problem is formalized with a MDP, $\sM$, defined as a tuple of state space, action space, transition model, reward model, and discount factor~\cite{puterman_1994}:
\begin{align}
    \label{eq:mdp}
    \sM &= \langle \sS,\sA,\fP,\fR,\gamma \rangle.
\end{align}

\begin{itemize}[leftmargin=*]
\item The state space, $\sS$, is defined as the set of map cells shown in Fig.~\ref{fig:gridworld_smallscale_statespace}, where the state of the $i\th$-taxi, $s^i_t$ is the cell index that contains that taxi's position. The number of cells in the environment is denoted by the cardinality of the set, $|\sS|$. 

\item The action space, $\sA$, is defined as a movement between map cells for for taxi $i$. The taxi on dispatch has 5 actions: $a^i_t \in \sA = \{\mathrm{stay},\mathrm{right},\mathrm{up},\mathrm{left},\mathrm{down}\}$, defined with respect to its current map cell in $\sS$. 

\item The transition function, $\fP: \sS \times \sA \times \sS \rightarrow \mathbbm{P}$ is defined as $\fP(s^i_t,a^i_t,s^i_{t+1}) = \pP(s^i_{t+1}|s^i_t,a^i_t)$. We define $\fP$ with a deterministic, cell-based dynamical model, $\ff$:
\begin{align}
    \label{eq:deterministic_transition}
    s^i_{t+1} &= \ff(s^i_t,a^i_t) \text{ where } \fP(s^i_t,a^i_t,s^i_{t+1}) = 1.
\end{align}
If the next state is valid, the cell-based dynamical model moves the taxi from the initial state, $s^i_t$ to the neighboring state $s^i_{t+1}$ according to its action. If the next state is not valid, the dynamical model returns the initial state. 

\item The reward function $\fR_t: \sS \times \sA \rightarrow \eR$ is defined to be the negative of the expected customer waiting time and is estimated from reward samples; reward is the negative of the time it takes the taxi to go from its current position to the dispatch position defined by the cell-action, and then from that position to the customer request. The reward has a subscript $t$ because we assume the reward changes over time due to the changing customer distribution. We equivalently write the $\fR_t$ function as a vector of state-action pairs, $\fR_t(s,a) = \vR_t[s|\sA| + i(a)]$ and $\vR_t \in \eR^{n_q}$, where $n_q = |\sS||\sA|$ and $i(a)$ denotes the action's index. Given a customer request, $\vc^k$, we compute a sample of the reward function, $\fr^i_t$ that can be used to estimate the underlying reward, $\vR_t$ according to an observation model:
\begin{align}
    \label{eq:reward_sample_def}
    \fr^i_t(s^i_t,a^i_t) &= -( \feta(\vp^i_t,\vu^i_t) + \feta(\vu^i_t,\vp^{k,p}) ) \\ 
    \label{eq:reward_sample}
    \vr^i_t &= \mH^i_t \vR_t + \vv^i_t
\end{align}
Recall that $\vp^i_t$ is the position of the $i\th$-taxi, $\vu^i_t$ is the dispatch desired position, and $\vp^{k,p}$ is the $k\th$-customer pickup position. 
Also, $\feta$ is the estimated time-of-arrival function that accepts position vectors, returns a scalar time value, and is specified in Appendix~\ref{appendix:fleet_simulator}. It is parameterized by the average taxi velocity $\overline{v}_\mathrm{taxi}$.
The measurement noise is sampled from a normal distribution with variance $\varsigma$, $\vv^i_t~\sim~\mathcal{N}(0,\varsigma I)$. 
Finally, the cell-based observation model, $\mH^i_t \in \eR^{n_q \times n_q}$, is a binary diagonal matrix with unity elements at the state-actions pairs where the customer request $\vc^k$ contains information of the corresponding state-action pair, and $0$ otherwise.

\item Note that $\gamma$ is the discount rate of the system. This parameter determines the trade-off between greedy and long-term optimal behavior. 
    
\end{itemize}

\section{Algorithm Description and Analysis: H-TD\textsuperscript{2}}
\label{sec:algorithm}
We describe the details of H-TD\textsuperscript{2} in this section, defining the exact and approximate policy estimation, the hybrid switching behavior, and the game theoretic task assignment. The overview of the method is given in Algorithm~\ref{algo:htd2}.

\begin{algorithm}
    \label{algo:htd2}
    \SetAlgoLined
    \textbf{input}: set of total, free, and servicing taxis: $\sI,\sI_f,\sI_s$ \\
    \textbf{output}: action profile for free taxis, $\vu_t$ \\
    \tcc{Hybrid Temporal Difference}
    \For{$\forall i \in \sI$}{
        \eIf{$\delta_e > \delta_d$~\eqref{eq:error}}{
            slow update $\vQ^i_t$ with aggregated global information~\eqref{eq:bellman} at the central node\;
            }{
            fast update $\vQ^i_t$ with local information~\eqref{eq:dtd} at each taxi\;
            }
        }
    \tcc{Game Theoretic Task Assignment}
    randomly initialize cell-based action profile $\tempA_t$\;
    \While{$\tempA_t$ not converged}{
        randomly pick $i\in\sI_f$ that has not converged\;
        consider current action, $a^i_t$\;
        propose random action $a^{i'}_t$\;
        compute marginal utility, $J$ with $\vQ^i_t$~\eqref{eq:marginal_utility}\;
        stochastically assign action with $J$~\eqref{eq:blll}\;
        check $i\th$-taxi action convergence\;
        }
    Convert cell-based actions $a^i_t$ to position vectors $\vu^i_t$\;
\caption{H-TD\textsuperscript{2} at timestep $t$}
\end{algorithm}

\subsection{Centralized $\fQ$-value Computation}
We present the idealized Bellman solution to the cell-based decision making problem specified in~\eqref{eq:mdp}. The solution is a policy function that maps states to an action that maximizes the discounted reward over time and can be represented as a value function, as shown in Fig.~\ref{fig:gridworld_smallscale_statespace}, or an action-value function known as $\fQ\bellman_t$-values. We use the latter and use the superscript $b$ notation to denote the policy that is synthesized with a Bellman iteration method. We adopt the conventional optimal $\fQ\bellman_t$-value function as follows:
\begin{align}
    \fQ\bellman_t(s_t,a_t) &= \mathbbm{E}_{ s \sim \fP(s_t^\prime| s_t, a_t)} [\fR_t + \gamma \mathbbm{E}_{a_t^\prime \sim \pi\bellman} \fQ\bellman(s_t^\prime,a_t^\prime)] 
\end{align}
We specify this general formulation with some assumptions. First, we use a finite-dimensional tabular $\fQ\bellman_t$ and write the $\fQ\bellman_t$ function as a vector of state-action pairs, $\vQ\bellman_t \in \mathbbm{R}^{n_q}$, as done with the reward in Sec.~\ref{sec:dispatch_mdp}. Next, we apply the deterministic transition function, $\fP(s^i_t,a^i_t,s^i_{t+1})$, as specified in~\eqref{eq:deterministic_transition} to remove the outer expectation. We remove the inner expectation by specifying the policy $\pi\bellman$ to be a transition kernel matrix $\mF\bellman$ such that $\mF\bellman \vQ\bellman_t(s^i_t) = \max_{a^i_t} \fQ\bellman_t(s^i_t,a^i_t)$. Combining these, we rewrite a simplified expression for $\vQ\bellman_t$ as the fixed point of the Bellman operator $\oT$:
\begin{align}
    \label{eq:bellman}
    \vQ\bellman_t 
    &= \vR_t + \gamma \mF\bellman \vQ\bellman_t 
    = \oT \vQ\bellman_t
\end{align}
The Bellman iteration can be solved using conventional value iteration or policy iteration methods from batch customer data. For the purpose of this paper, we use a Modified Policy Iteration (MPI) method~\cite{puterman_1978} to solve line 5 of Algorithm~\ref{algo:htd2}. However, there are complications with implementing a pure Bellman approach, which we address in the next subsection. 

\subsection{Distributed Reward Estimation and $\vQ$-value Iteration}
\label{sec:dtd}
We present a policy approximation that overcomes the Bellman solution's practical limitations. The first issue with a pure Bellman solution is that in an online setting, the reward information is hidden a priori and received incrementally in samples, $\vr^i_t$. So the expectation of the reward is not immediately available. Furthermore, each taxi only has access to local information. To overcome these problems, we assume the hidden reward evolves as a random walk process and synthesize linear estimators for the hidden reward $\vR_t$: 
\begin{align}
    \label{eq:reward_dynamics}
    \vR_{t+1} &= \vR_t + \vw_t \\ 
    \label{eq:ctd_reward_estimator}
    \vR\td_{t+1} &= \vR\td_t + \sum_{j\in\sI} K^j_t (\vr^j_t - \mH^j_t \vR\td_t) \\ 
    \label{eq:dtd_reward_estimator}
    \vR^i_{t+1} &= \vR^i_t + \sum_{j\in\sI} \mA^{ij}_t (\vr^j_t - \vR^i_t) 
\end{align}
Recall $\vr^i_t$ is the reward sample defined in~\eqref{eq:reward_sample}. 
Also $\vR\td_t,\vR^i_t \in \mathbbm{R}^{n_q}$ are centralized and distributed reward estimators that will be used in the upcoming temporal difference learning, where the superscript $\mathrm{c}$ denotes a centralized quantity. 
The $\mH^i_t$ matrix is the $i\th$-taxi's measurement model defined in~\eqref{eq:reward_sample} and the $K^i_t$ matrix are the corresponding estimator gains. The process noise, $\vw_t~\sim~\mathcal{N}(0,\varepsilon I)$ is sampled from a normal distribution where the parameter $\varepsilon$ is computed offline from training data. 
The row-stochastic adjacency block matrix, $A_t$, specifies the local information available to each agent and is defined as:
\begin{alignat}{2}
    \label{eq:adjacency}
    \mA^{ij}_t &= \mB^{ij}_t / \sum_{j=1}^{n_i} \mB^{ij}_t, \text{ where } 
    \mB^{ij}_t = 
    \begin{cases}
        \mK^j_t \mH^j_t & j \in \sI^i_t \\ 
        \mathbb{0}_{n_q \times n_q} & \mathrm{else} 
    \end{cases} \\ 
    \label{eq:neighbors}
    \sI^i_t &= \{ j \in \sI \ \ | \ \ \|\vp^i_t - \vp^j_t\| < R_\mathrm{comm}\}.
\end{alignat}
with diagonal block matrices $\mA^{ij}_t, \mB^{ij}_t \in \eR^{n_q \times n_q}$ and full matrices $\mA_t, \mB_t \in \eR^{n_i n_q \times n_i n_q}$. The i$\th$ agent constructs the $\mB^{ij}_t$ matrices from the gain and measurement matrices shared by its neighbors $j\in\sI^i_t$. The local observation is parameterized by the radius of communication, $R_\mathrm{comm}$. 

The second issue with the Bellman approach is an intrinsic drawback that at higher state/action dimensions the Bellman-iteration calculation becomes computationally-expensive and cannot be quickly evaluated online. Instead, temporal difference learning~\cite{szepesvari_2010} can be used as an approximate method to estimate $\vQ$-values online using $\vR\td_t$:
\begin{align}
    \label{eq:ctd} \vQ\td_{t+1} 
    &= \vQ\td_t + \alpha (\vR\td_t + \gamma \mF\td \vQ\td_t - \vQ\td_t)
\end{align}
where $\alpha \in \eR$ is the system learning rate and $\mF\td$ is the transition kernel for this policy. 

This formulation requires a central node to collect the data, compute a policy, and broadcast the new information at every timestep, scaling the computation complexity, bandwidth, and network delay with the number of taxis.
To address this limitation, we introduce a distributed algorithm using communication between the taxis, and propose a policy update computed at each taxi using only local information:
\begin{align}
    \label{eq:dtd}
    \vQ^i_{t+1} &= \vQ^i_t + \alpha( \vR^i_t + \gamma \mF^i \vQ^i_t - \vQ^i_t)
\end{align}
where this temporal difference~\eqref{eq:dtd}, with the distributed reward estimation~\eqref{eq:dtd_reward_estimator} defines line 7 of Algorithm~\ref{algo:htd2}. 

Using the approximate temporal difference method and estimating with only local information hurts the quality of the final policy used by each taxi. We derive the upper bound of the negative effect of these approximations through a regret bound analysis, comparing the policy synthesized with the proposed algorithm~\eqref{eq:dtd} and the Bellman-optimal solution~\eqref{eq:bellman}. 

\begin{theorem}
\label{theorem:i_to_b}
The expected distance between an arbitrary taxi $\vQ$-value estimates, $\vQ^i_t$, and the Bellman-optimal solution $\vQ\bellman_t$ is upper bounded by: 
\begin{align}
    &\mathbbm{E} \| \vQ^i_t - \vQ\bellman_t \|_2 \leq 
    \frac{2 \sqrt{n_q (\varepsilon + \varsigma)} }{(1-\gamma)(1-\sqrt{1 - \lambda_\mathrm{min}(\sum_{j\in\sI} \mA^{ij}_t)})}
\end{align}
where $\lambda_\mathrm{min}(.)$ denotes the smallest eigenvalue of a matrix. 
\end{theorem}

\begin{IEEEproof}
First, we write the distributed iteration as an application of the Bellman operator on the previous timestep with a disturbance. Then we solve for the disturbance to derive the final bound.  

\emph{Step 1}:
Consider~\eqref{eq:dtd} and add and subtract $\alpha \vR_t$:
\begin{align*}
    \vQ^i_{t+1} 
    &= \vQ^i_t + \alpha( \vR^i_t + \gamma \mF^i \vQ^i_t - \vQ^i_t) + \alpha \vR_t - \alpha \vR_t \\ 
    &= \vQ^i_t + \alpha( \vR_t + \gamma \mF^i \vQ^i_t - \vQ^i_t) + \alpha (\vR^i_t - \vR_t) \\ 
    &= \vQ^i_t + \alpha( \oT \vQ^i_t - \vQ^i_t) + \alpha \ve^i_t 
\end{align*}
where $\ve^i_t = \vR^i_t - \vR_t$. The system is contracting at rate $1-\alpha (1-\gamma)$, implying the system geometrically converges to an equilibrium about $\oT \vQ^i_t = \vQ^i_t$. In addition, from Banach's fixed point theorem~\cite{stuart_1996}, $\oT$ contracts to a unique fixed point, $\vQ\bellman_t$. Applying Discrete Gronwall's lemma~\cite{stuart_1996}: 
\begin{align*}
    \| \vQ^i_t - \vQ\bellman_t \| \leq \frac{\|\ve^i_t\|}{1-\gamma}
\end{align*}
Note that the decaying initial condition term does not appear because we assume that the policy estimate is initialized with the Bellman solution, i.e. $\vQ^i_0 = \vQ\bellman_0$. 

\emph{Step 2}:
Here we need to bound the value $\|\ve^i_t\|$, i.e. the error between the estimated reward and the true reward. We write the dynamics of the error vector $\ve^i_t$ by subtracting~\eqref{eq:reward_dynamics} from~\eqref{eq:dtd_reward_estimator}:
\begin{align*}
    \ve^i_{t+1} &= (I - \sum_{j\in\sI} A^{ij}_t) \ve^i_t + \vd_t
\end{align*}
where $\vd_t = \vw_t + \sum_{j\in\sI} A^{ij}_t \vv^j_t$. By applying Weyl's interlacing eigenvalue theorem~\cite{horn_2012}, we prove that the $i\th$ system is contracting at rate lower bounded by $\lambda^i_t = 1 - \lambda_\mathrm{min}(\sum_{j\in\sI} A^{ij}_t)$. 

We rewrite the disturbance as the product of an input matrix, $\mM$, and the stacked noise, $\vz_t\sim\mathcal{N}(0,\mW)$:
\begin{align*}
    \vd_t &= \mM_t \vz_t 
\end{align*}
where $\vz_t = \begin{bmatrix} \vw_t; \vv^1_t; \hdots; \vv^{n_i}_t \end{bmatrix}$, $\mM_t = \begin{bmatrix} I_{n_q}, \mA^{i,1}_t, \hdots, \mA^{i,n_i}_t \end{bmatrix}$ and $\mW = \mathrm{blkdiag}(\varepsilon I_{n_q}, \varsigma I_{n_q}, ..., \varsigma I_{n_q})$. 

By application of the convergence theorem of discrete stochastic contracting systems~\cite{Lohmiller_1998,Tsukamoto_2020}, the expected error of a single agent is upper bounded by: 
\begin{align*}
    \mathbbm{E} \|\ve^i_t \| &\leq \frac{2 \sqrt{C}}{1-\sqrt{1 - \lambda_\mathrm{min}(\sum_{j\in\sI} \mA^{ij}_t)}} \\ 
    C &= \trace(\mM_t^T \mM_t \mW)
\end{align*}
It remains to calculate the value $C$:
\begin{align*}
    C &= \varepsilon \trace(I_{n_q}) + \varsigma \sum_{j\in\sI} \trace \left( (\mA^{ij}_t)^T \mA^{ij}_t \right) \leq n_q (\varepsilon + \varsigma) 
\end{align*}
where we use the linearity of the trace operation to move it outside of the sum, then we use the non-negativity and row-stochasticity of $\mA_t$ to bound $\sum_{j\in\sI} (\mA^{ij}_t)^T \mA^{ij}_t \leq I_{n_q}$.  
The final result is found by plugging the result from Step 2 into the result from Step 1.  
\end{IEEEproof}

\begin{remark}
    The regret bound is driven by the contraction rate of the system $\lambda^i_t$, a combined graph and observeability quantity. Intuitively, this corresponds to a non-zero value when taxi $i$ and its neighbors taxis $j$ can measure the entire state-action vector. We can also consider a batch measurement over a time interval, $n_T$ and an average contraction rate, $\overline{\lambda^i_t}$. This time interval approach exists in the multi-agent adaptive control literature, where $\overline{\lambda^i_t}>0$ is analogous to an excitation level in the Collective Persistency of Excitation condition~\cite{wensing_2018}. 
\end{remark}

\begin{remark}
    The proposed online method is used to estimate the optimal policy in a dynamic environment, i.e. the reward model, $\vR_t$ is time-dependent. In this case, the optimal policy is non-stationary, i.e. $\vQ\bellman_{t+1} \neq \vQ\bellman_t$. In order to guarantee the convergence of the TD-algorithm, we require that there exists a timescale separation between the convergence of the TD-algorithm and the dynamics of $\vQ\bellman$:
    \begin{align*}
        \| \vQ\bellman_{t+1} - \vQ\bellman_t\|  \ll (1-\gamma)(1-\sqrt{1-\lambda_\mathrm{min}(\sum \mA^{ij}_t)}) 
    \end{align*}
\end{remark}

\begin{remark}
The adjacency matrix $\mA_t$ dictates that each agent takes a convex combination of the neighboring measurements. Further, the estimation gain matrices, $\mK^i_t$ are chosen with a Distributed Kalman Information Filter, whose proof of optimality with respect to mean-squared-error can be found in~\cite{saber_2009,bandyopadhyay_2018}. Thus, the agents weigh the neighboring measurements appropriately. 
\end{remark}

\subsection{Hybrid Temporal Difference Algorithm}

We define the switching condition in line 4 of Algorithm~\ref{algo:htd2} with two parameters, $\delta_e$, the estimated error in the system, and $\delta_d$, the user specified desired error in the system. 

\begin{proposition}
    If we define: 
    \begin{align}
        \label{eq:error}
        \delta_e &= \frac{2\sqrt{n_q (\varepsilon + \varsigma)}}{(1-\gamma)(1-\sqrt{1 - \lambda_\mathrm{min}(\sum_{j\in\sI} \mA^{ij}_t)})}
    \end{align}
    the expected policy sub-optimality will be bounded by $\delta_d$. 
\end{proposition}

Nominally, the system evolves with the distributed temporal difference method~\eqref{eq:dtd}, computing $\delta_e$ at each timestep. Each agent is able to compute this value because $\varepsilon,\varsigma$, and $\gamma$ are known system parameters and each agent keeps track of its own $\lambda_\mathrm{min}(\sum_{j\in\sI} \mA^{ij}_t)$ values. Applying the result from Theorem~\ref{theorem:i_to_b}, the expected policy suboptimality is identically $\delta_e$, so, if $\delta_e$ exceeds the desired error, $\delta_d$, the desired sub-optimality is violated. However, if this condition occurs at time $t$, the system resets all taxis with a central policy update~\eqref{eq:bellman}, i.e. $\vQ^i_t = \vQ\bellman_t$, $\forall i \in \sI$. Therefore, the H-TD\textsuperscript{2} algorithm maintains the distance between the estimated policy and a true regret policy to user specification. In effect, $\delta_d$ controls the trade-off of computational expense to policy sub-optimality, where $\delta_d \rightarrow 0$ produces a solution with no regret but maximum computational effort and $\delta_d \rightarrow \infty$ produces a solution with potentially infinite regret with little computational effort. 

\subsection{Game Theoretic Task Assignment}
In this section, we propose a game-theoretic task assignment to coordinate the taxis according to the $\vQ^i_t$-values estimated in Sec.~\ref{sec:dtd}. The $\vQ^i_t$ policy does not account for the actions of the other taxis and, without additional coordination, the taxis would behave greedily by all going to the highest value cell, increasing the overall customer waiting time. To avoid this behavior, we design a potential game and a local action profile iteration to maximize each agent's marginal utility. This is implemented in Algorithm~\ref{algo:htd2}, Lines 10-18.

First, we introduce a global action profile, $\tempA_t$ and a local action profile for the $i\th$ taxi, $\tempA^i_t$: 
\begin{alignat}{2}
    &\tempA_t = \{ a^j_t \ \ | \ \forall j \in \sI \}, \text{ and }
    &\tempA^i_t = \{ a^j_t \ \ | \ \forall j \in \sI^i_t \}
\end{alignat}
where the $j\th$-neighbor taxi communicates its action, $a^j_t$, to the $i\th$-taxi, where $a^j_t$ is defined in~\eqref{eq:mdp}.

Next, we introduce the current global fleet distribution $\sOmega_t$, i.e. the number of taxis in each cell, as a function of the action profile: 
\begin{align}
    \sOmega_t(s,\tempA_t) &= \frac{1}{n_i} \sum_{j\in\sI} \I(s = \ff(s^j_t,a^j_t))
\end{align}
where $\I$ denotes the indicator function and $\ff$ is the dynamics model specified in~\eqref{eq:mdp}. The local fleet distribution, $\sOmega^i_t$, is found with the same calculation but summing only over the neighboring agents, $j\in\sI^i_t$. By defining the radius of communication as $R_\mathrm{comm} = 3 \ds$ where $\ds$ is the length of a cell in the environment, we guarantee that the $i\th$ taxi can always calculate the fleet distribution in neighboring cells, $\sS^i$, within one action of the current cell of the $i\th$ taxi.

Next, we describe the desired fleet distribution using the $\vQ$-values as computed in~\eqref{eq:dtd_reward_estimator}. Consider the following Boltzmann exploration strategy~\cite{szepesvari_2010} strategy to synthesize a desired distribution, $\ffpi$, from the $\vQ$-value estimates:
\begin{align}
    \ffpi(s,\fQ^i_t) &= \frac{\exp \left( \beta \max_{a\in\sA} \fQ^i_t(s,a) \right)}{ \sum_{a'\in \sA} \exp \left( \beta \fQ^i_t(s,a') \right) }
\end{align}
Recall that $\sA$ are the local actions available to each taxi defined in~\eqref{eq:mdp} and $\beta \in \eR$ is an exploration/exploitation design constant. For simplicity of notation, we have written the action-values in its functional form, $\fQ^i_t$.

The goal of the game theoretic task assignment is to find an action profile, $\tempA_t$ through local iteration methods that minimizes the distribution distance between the current distribution, $\sOmega_t$ and the desired distribution $\ffpi$. We describe a potential and noncooperative game meaning that the taxis will try to converge to a Nash equilibrium with a high potential function value. The global and marginal potential functions, $\fPhi$ and $\fJ$ are defined as follows:
\begin{align}
    \label{eq:game}
    \fPhi(\tempA_t) &= -\sum_{s\in\sS} (\ffpi(s,\fQ^i_t) - \sOmega(s,\tempA_t))^2 \\
    \label{eq:marginal_utility}
    \fJ(\tempA^i_t) &= -\sum_{s\in\sS^i} (\ffpi(s,\fQ^i_t) - \sOmega(s,\tempA^i_t))^2 
\end{align}
For the calculation of $\fJ$, we only require the indices that correspond to neighboring cells of the current cell of the $i\th$-taxi, thereby permitting a local calculation. 
\begin{remark}
    The game's utility function, $\fPhi(\tempA_t)$ has an analogy to sample-based planners if each taxi in the fleet is considered as a sampled action of a stochastic policy, $\ffpi(s,\vQ^i_t)$. This choice of utility function is interesting because it could be the utility function chosen by a centralized algorithm but we can maximize it with local calculations through $\fJ$. 
\end{remark}
\begin{remark}
Note that $J$ is indeed the marginal contribution on the global potential function: 
\begin{align*}
    &\fPhi(\tempA'_t) - \Phi(\tempA_t) = \fJ(\tempA^{i'}_t) - \fJ(\tempA^{i}_t)
\end{align*}
where $\tempA^{'}_t = \{ a^j_t | \ \forall j \in \sI / \{i\} \} \cup \{a^{i'}_t\}$ is the global alternative action set. 
\end{remark}

We use a game-theoretic reinforcement learning technique, binary log-linear learning~\cite{marden_2012} to iterate to an action set $\tempA_t$, shown in line 10-19 of Algorithm~\ref{algo:htd2}. At each timestep, $t$, the action set, $\tempA_t$ is randomly initialized. While all other taxi's actions are held, a randomly selected $i\th$-taxi chooses between the previously held action, $a^i_t$, and an alternate action $a^{i'}_t$ with probability $p^i_t(\tempA^i_t, \tempA^{i'}_t)$: 
\begin{align}
    \label{eq:blll}
    p^i_t(\tempA^i_t, \tempA^{i'}_t) &= 
    \frac{\exp{(\fJ(\tempA^i_t)/\tau)}}{\exp{(\fJ(\tempA^i_t)/\tau)} + \exp{(\fJ(\tempA^{i'}_t)/\tau)}} 
\end{align}
where $\tempA^{i'}_t = \{ a^j_t | \ \forall j \in \sI^i_t / \{i\} \} \cup \{a^{i'}_t\}$ is the local alternative action set. The coefficient $\tau \in \eR_{>0}$ is a design parameter specifying how likely taxi $i$ chooses a sub-optimal action, to specify the trade-off between exploration and exploitation. The action set is chosen once the iteration has converged, completing the game-theoretic task assignment. Then, the cell-based action $a^i_t$ is converted to a dispatch vector $\vu^i_t$, where $\vu^i_t$ is a randomly sampled position vector in the cell after the dispatch action is taken.  

\section{Numerical Experiments}
\label{sec:numerical_experiments}
\subsection{Baseline and Variants}
We compare our H-TD$^2$ solution with a receding horizon control (RHC) baseline dispatch algorithm that is similar to the baseline in~\cite{oda_2018}. For this section, we use independent notation from the rest of the paper, matching the notation in~\cite{oda_2018}. The RHC dispatch algorithm is formulated as the following linear program:
\begin{equation}
\begin{alignedat}{2}
	\vu^* &= 
	\argmax \sum_{t=t_0}^{t_0+t_\mathrm{rhc}} \gamma^{t-t_0} \sum_{i=1}^{M} \min(\overline{w}_{t,i} - \vx_{t,i},0) \quad \text{ s.t. } \\
    & \quad \sum_{j=1}^{M} u_{ij,t} = \vx_{t+1,i}, \quad \sum_{i=1}^{M} u_{ij,t} = \vx_{t,i}  \\ 
    & \quad u_{ij,t} = 0 \quad \forall j \notin \sS^i, \quad \vx_{t_0,i} = \mX_{0,i}  
\end{alignedat}
\end{equation}
where the state variable, $\vx_t \in \Z^{|S|}$, is the number of free taxis in each cell, the control variable $u_{ij,t}$ is the number of taxis moving from cell $i$ into cell $j$ at time $t$ and $t_\mathrm{rhc}$ is the RHC planning horizon. The first two constraints are conservation constraints: (i) the number of taxis in cell $i$ is the number of taxis moving into cell $i$, and (ii) the number of taxis moving from cell $i$ is equal to the number of taxis previously in cell $i$. 
The third constraint is that each taxi can only move to a neighboring cell. The fourth constraint is the initial condition. For large fleets, a proper assumption from~\cite{oda_2018} is to relax $\vu_{ij,t}$ from integers to real numbers, resulting in a linear program in $\vu_{ij,t}$. The reward is the difference of customer demand and taxi supply, where $\overline{w}_{t,i}$ represents the expected customer demand and computed from training data similar to that proposed in~\cite{miao_2016,oda_2018}. 
The baseline is chosen to demonstrate the advantage of fast adaption over learned prediction: static prediction methods (either explicit in model-based or implicit in model-free) are vulnerable to events occurring outside of the training domain.


We study the effect of each component of the policy estimation algorithm by comparing variants of the policy: \emph{Centralized Temporal Difference} (C-TD), \emph{Distributed Temporal Difference} (D-TD), and \emph{Bellman-optimal}. All of these variants control the fleet with binary log-linear learning~\eqref{eq:blll}, but they differ in how $\vQ$ is synthesized: C-TD uses~\eqref{eq:ctd_reward_estimator} and~\eqref{eq:ctd}, D-TD uses \eqref{eq:dtd_reward_estimator} and~\eqref{eq:dtd}, and the Bellman-optimal policy is synthesized with~\eqref{eq:bellman}. Equivalently, the D-TD and Bellman-optimal algorithms can be interpreted as the limiting behavior of H-TD\textsuperscript{2} corresponding to the respective cases where $\delta_d = \infty$ and $\delta_d = 0$.  

\subsection{Customer Demand Datasets}
We consider two datasets of customer requests: a synthetic dataset for a Gridworld environment and the real customer taxi dataset from the city of Chicago~\cite{chicago_data}. Recall each customer request is defined as follows: $\vc^k = [t^k_r,t^k_d,\vp^{k,p},\vp^{k,d}]$. 

The synthetic dataset is generated as follows: At each timestep, $t \in [t_0, t_0+\Delta_t]$, the customer request model is sampled $n_c$ times, where $n_c$ is the number of customers per timestep of the simulation. The time of the request, $t^k_r$ is uniformly randomly sampled in the timestep. The pickup location, $\vp^{k,p}$, is found by sampling a 2-dimensional Gaussian Mixture Model (GMM), where translating Gaussian distributions capture the underlying dynamic customer demand. The dropoff location, $\vp^{k,d}$ is a randomly sampled position in the map, and the duration of the trip, $t^k_d$ is given by $\feta(\vp^{k,p},\vp^{k,d})$. The GMM model is parameterized by the number of Gaussian distributions, $n_G$, the speed of the distributions, $v_G$, the variance, $\sigma_G$, and the initial position, and unit velocity vector of the centroid for each distribution. 

The real customer dataset is taken from the city taxi dataset of Chicago~\cite{chicago_data} filtered by start and end timestamp. The Socrata API permits importing raw data in the $\vc^k$ format, where the location data is specified in longitude, latitude coordinates. For this experiment, we load a city map of Chicago as a shapefile and perform minor geometric processing with the Shapely Python toolbox.

\subsection{Results}

\textit{Gridworld Simulations:} We present variants of our H-TD\textsuperscript{2} algorithm against a RHC baseline in a Gridworld environment as shown in Fig.~\ref{fig:gridworld_smallscale_statespace}. This flexible, synthetic environment permits us to test on a range of system parameters. 

First, we introduce the algorithms in a small-scale simulation. We synthesize a dataset with parameters: $n_c=5$, $n_G=2$, $v_G=0.02625$, $\sigma_G = 0.014$, and randomly initialize the mean and direction of the distributions. Our simulation parameters are: $n_i = 100$, $|\sS| = 85$, $\gamma = 0.9$, $\alpha = 0.75$, $n_T = 10$, $\varsigma = 0.014$, $\varepsilon = 0.0187$, $\delta_d = 0.025 \|\vQ\bellman_0\|$, $\beta = 150$, $\overline{v}_\mathrm{taxi} = 0.125$, $\tau = 0.0001$, and $t_\mathrm{RHC} = 10$. We run this experiment for each of the algorithms for $5$ trials. This experiment is shown in Fig.~\ref{fig:gridworld_smallscale_statespace}, modified with $n_i =1000$. 

\begin{figure}
    \centering
    \includegraphics[width=0.85\linewidth,trim={0.5cm 0.25cm 1.25cm 1.25cm},clip]{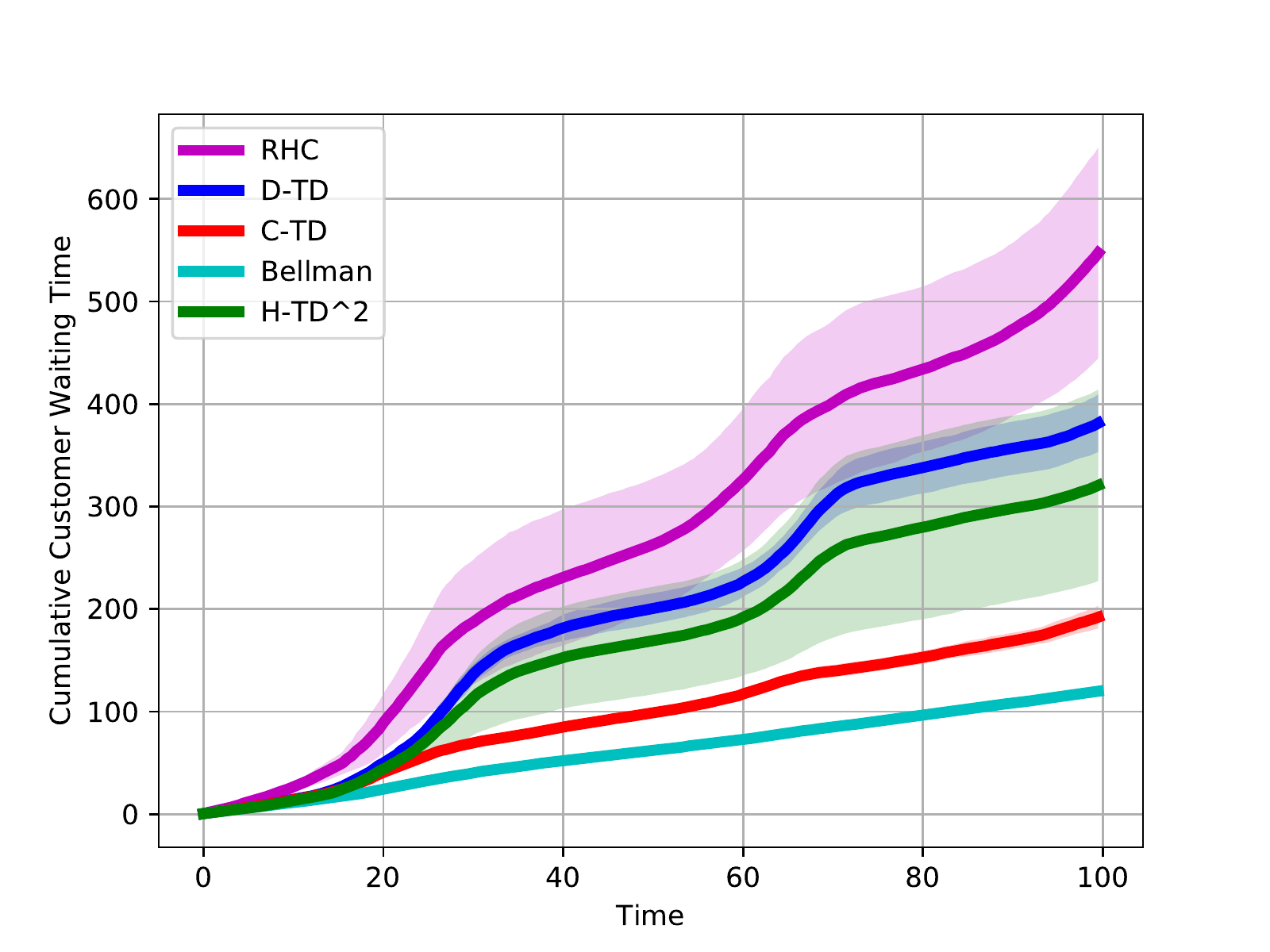}    
    \caption{Cumulative customer waiting time for different algorithms in the small-scale Gridworld environment. The algorithms behave as expected: in descending order of performance, Bellman, centralized temporal difference, H-TD\textsuperscript{2}, distributed temporal difference, followed by the receding horizon control baseline. At the cost of computational effort, the user can tune the performance of H-TD\textsuperscript{2} between the distributed temporal difference and Bellman-optimal solution by changing $\delta_d$. The simulation is run $5$ times, and the mean with standard deviations is visualized in the plot.}
    \label{fig:gridworld_smallscale_reward}
\end{figure}

Next, we collect statistics on the cumulative reward of each algorithm, and plot the results in Fig.~\ref{fig:gridworld_smallscale_reward}. The algorithms behave as expected: in descending order of performance, Bellman, centralized temporal difference, H-TD\textsuperscript{2}, distributed temporal difference, followed by the receding horizon control baseline. At the cost of computational effort, the user can tune the performance of H-TD\textsuperscript{2} between the distributed temporal difference and Bellman-optimal solution by changing $\delta_d$. The simulation is run $5$ times, and the mean with standard deviations is visualized in the plot.

\begin{figure}
    \centering
    \includegraphics[width=0.85\linewidth,trim={0.5cm 0.25cm 1.25cm 0.75cm},clip]{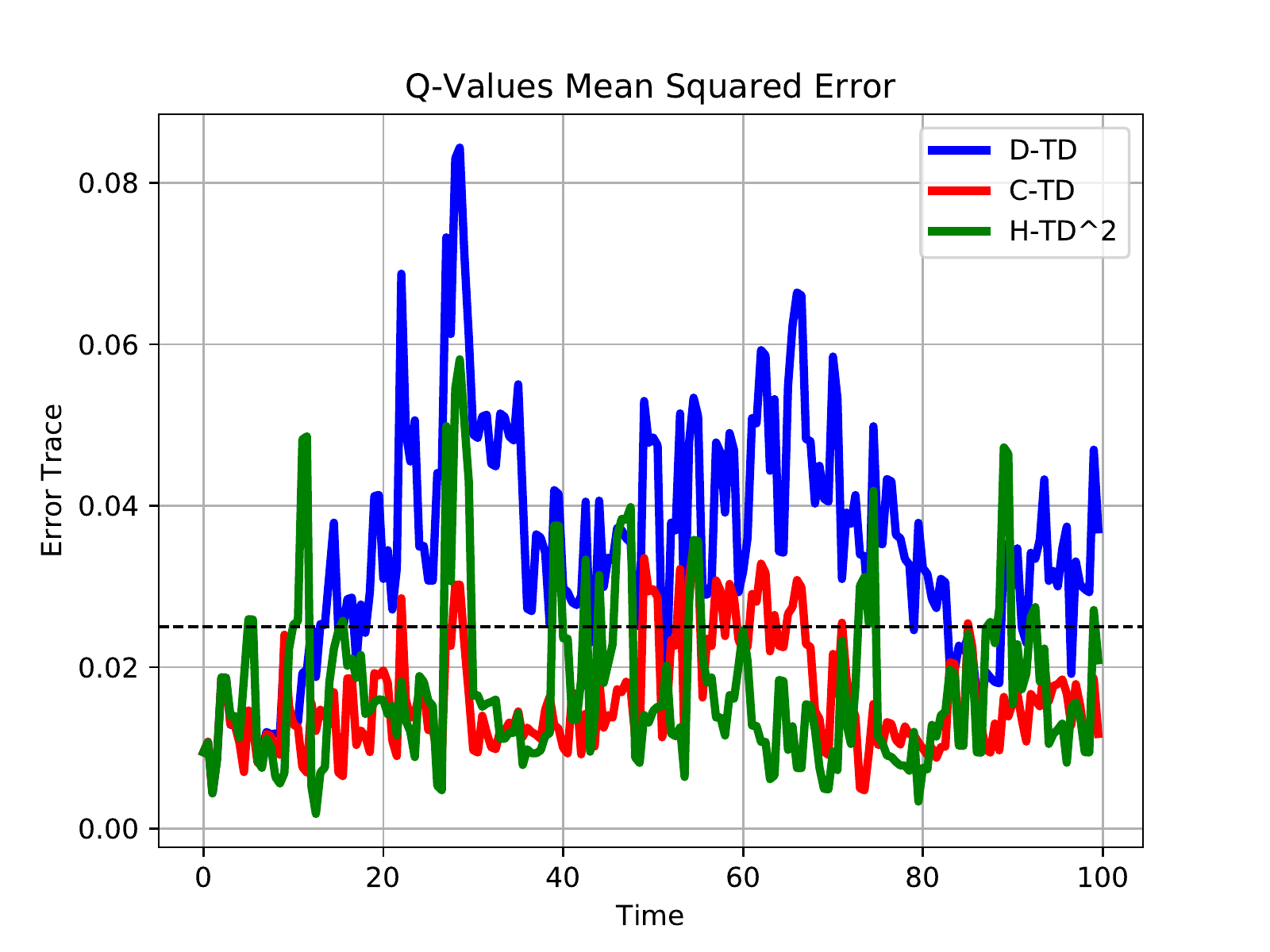}
    \caption{$\vQ$-value error trace for different algorithms with respect to the Bellman-optimal. Initially, the H-TD\textsuperscript{2} and distributed temporal difference algorithms behave identically, until the trigger condition is satisfied and the H-TD\textsuperscript{2} requests a global Bellman-optimal update, bringing the error to zero. The $\delta_d$ parameter is set to \SI{2.5}{\percent} of the norm of the Bellman solution and is shown with a dashed black horizontal line.}
    \label{fig:gridworld_smallscale_qmse}
\end{figure}

To explain the performance difference between the variants of our method, we show an error trace of the policy in Fig.~\ref{fig:gridworld_smallscale_qmse}. For a given policy $\vQ$, we calculate the error with respect to the Bellman solution: $e^{\vQ}_t = \| \vQ\bellman_t - \vQ_t \| / \| \vQ\bellman_t \|$. For this experiment, we set the $\delta_d$ parameter is set to \SI{2.5}{\percent} of the norm of the Bellman solution, indicated by the dashed horizontal line. Initially, the H-TD\textsuperscript{2} and distributed temporal difference (D-TD) algorithms behave identically, until the trigger condition is satisfied and the H-TD\textsuperscript{2} requests a global Bellman-optimal update, thereby bringing its error to zero. As expected, the centralized-temporal difference method, C-TD, generally tends to estimate the $\vQ$-values better than its distributed counterpart, D-TD. 

\begin{figure}
    \centering
    \includegraphics[width=0.85\linewidth,trim={0.25cm 0.45cm 0.25cm 0.45cm},clip]{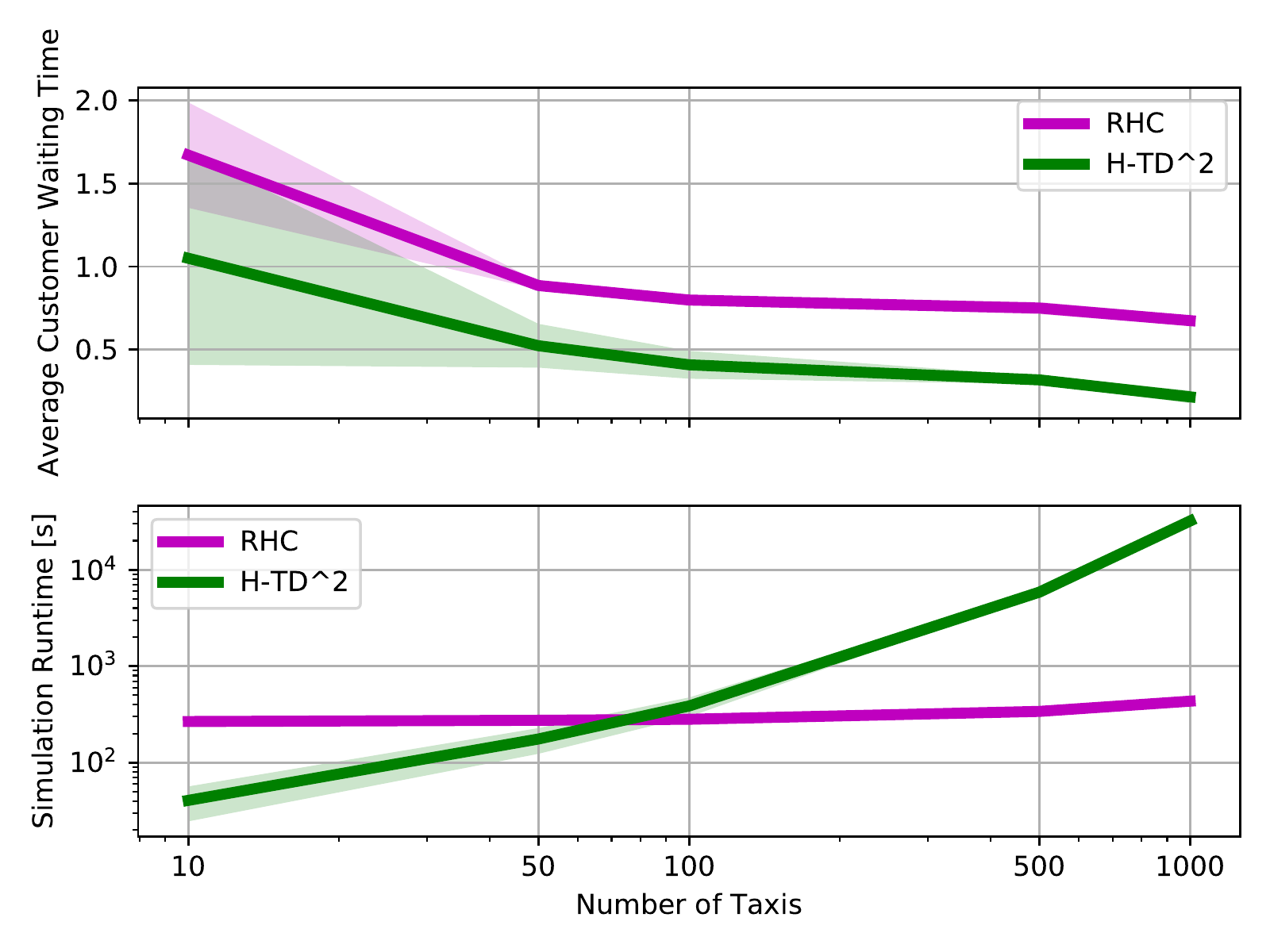}    
    \caption{Performance and scalability analysis of H-TD\textsuperscript{2} and RHC against number of taxis. In the top subplot, the average reward is shown across a variety of taxi density regimes, and the proposed algorithm outperforms a receding horizon control baseline by at least \SI{50}{\percent} in all taxi-density regimes. In the bottom subplot, the computational time is approximately linear with number of taxis (and taxi-density) across 3 orders of magnitude.}
    \label{fig:gridworld_scaleability_analysis}
\end{figure}

Next, we test the proposed algorithm and baseline's scalability and performance across a wide range of taxi-densities and plot the results in Fig.~\ref{fig:gridworld_scaleability_analysis}. We fix the parameters from the small-scale simulation and only change the number of taxis and number of customers, where we maintain the ratio $n_i / n_c = 10$. In the top figure, the average reward is shown across a variety of taxi density regimes, where the H-TD\textsuperscript{2} algorithm outperforms a RHC baseline by almost a factor of 2 in all taxi-density regimes. In the bottom figure, we show that the computational time is approximately linear with number of taxis (and taxi-density) across 3 orders of magnitude. The computational complexity of both RHC and H-TD\textsuperscript{2} scales with the spatial resolution of the simulation, and in practice, we limit the maximum number of cells to 200. 


\textit{Chicago City Simulations} We present the H-TD\textsuperscript{2} against an RHC baseline using real customer data from the city of Chicago public dataset~\cite{chicago_data}, in a Chicago map environment. We show that our algorithm outperforms the baseline in practical datasets and demonstrate that online algorithms are robust in irregular urban mobility events. 

\begin{figure}
    \centering
    \includegraphics[page=1,width=0.8\linewidth,trim={0.5cm 0.25cm 1.25cm 0.75cm},clip]{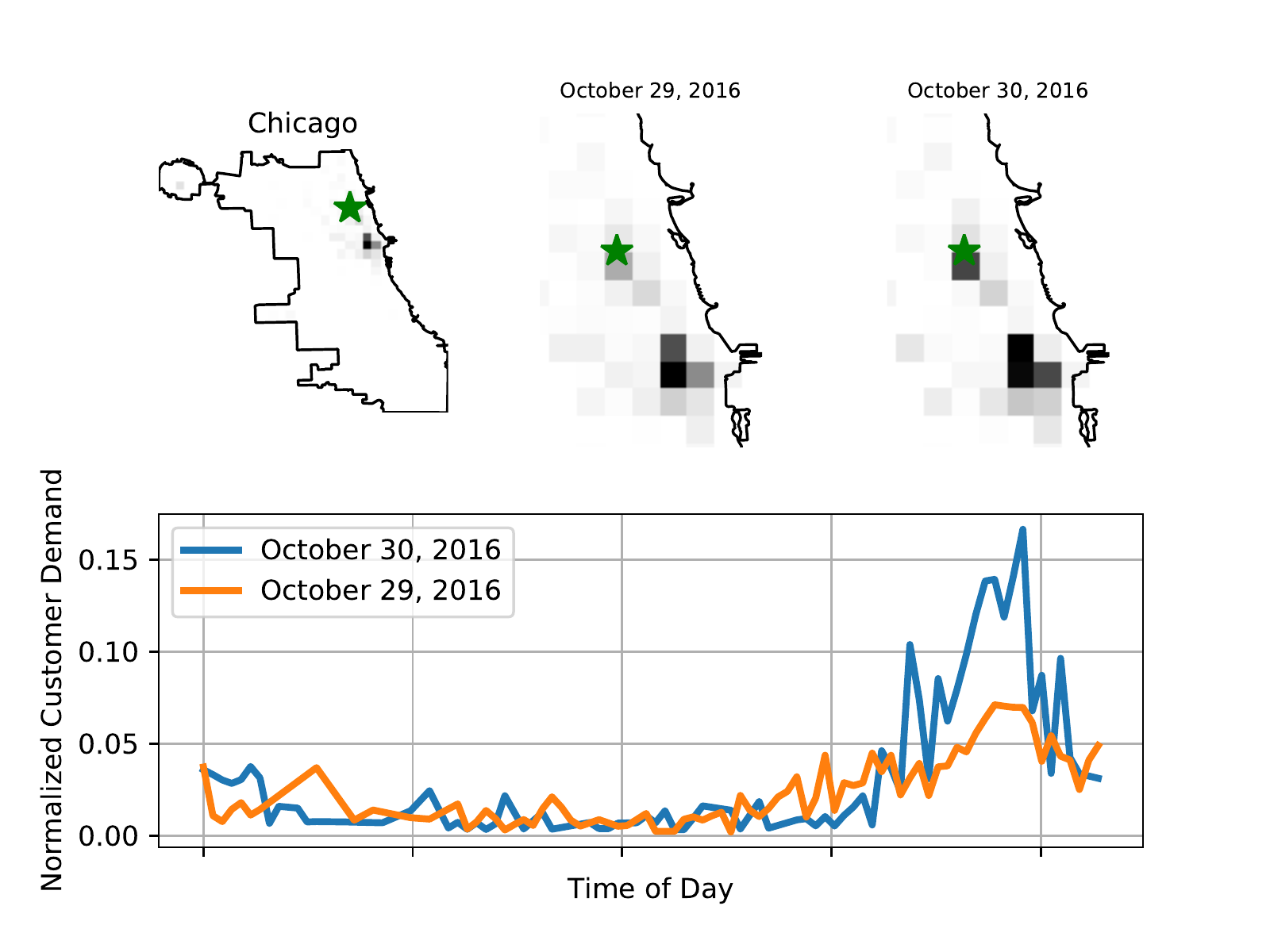}
    \caption{
    Chicago city taxi customer demand across an irregular event: Game 5 of the 2016 Baseball World Series. The map cells show the number of customer pickup requests, and the green star is Wrigley Field's location. Below, we plot the customer demand over time for the cell containing Wrigley Field to show that a reward model trained using data from the day does not accurately predict the future behavior.}
    \label{fig:chicago_heatmap}
\end{figure}

In Fig.~\ref{fig:chicago_heatmap}, we present the Chicago city taxi customer demand across an irregular event: Game 5 of the 2016 Baseball World Series. The map cells show the number of customer pickup requests, and the green star is Wrigley Field's (baseball stadium) location. Below, we plot the customer demand over time for the cell containing Wrigley Field. We show that a reward model trained using data from the day before would not accurately predict the behavior of the next day.

\begin{figure}
    \centering
    \includegraphics[page=2,width=0.8\linewidth,trim={0.125cm 0.25cm 1.25cm 0.75cm},clip]{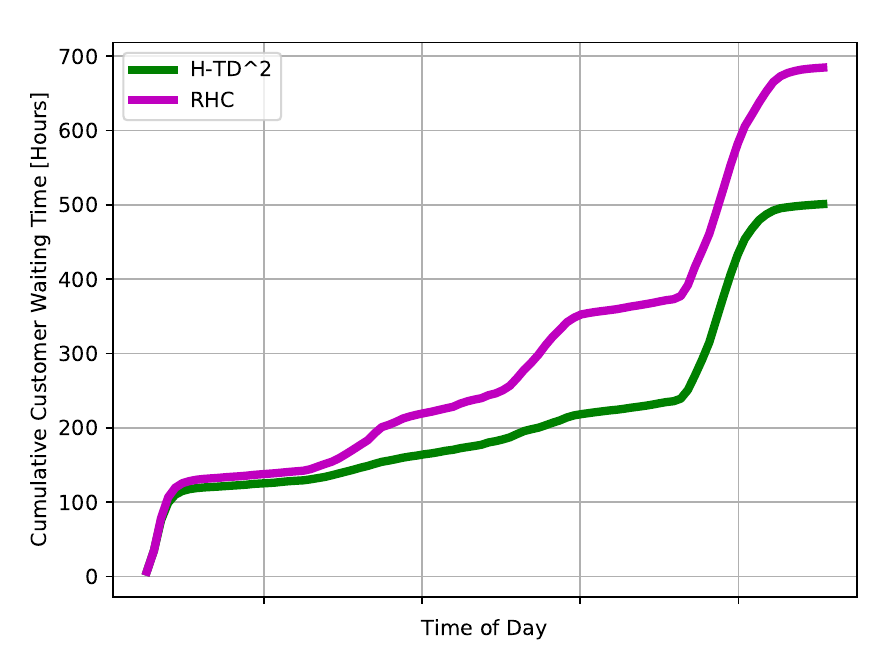}
    \caption{Cumulative customer waiting time for the H-TD\textsuperscript{2} and the RHC baseline in a Chicago city environment with with a fleet of $2,000$ taxis servicing $54,115$ real customer requests during the 2016 Major League Baseball World Series. The H-TD\textsuperscript{2} algorithm has a total customer waiting time of \SI{504} hours, an improvement of \SI{26}{\percent} over the RHC baseline. This result demonstrates the robustness of adaptive algorithms to irregular events.}
    \label{fig:chicago_reward}
\end{figure}

We evaluate the algorithms and plot the results in Fig.~\ref{fig:chicago_reward}. Our simulation parameters are: $n_i = 2000$, $|\sS| = 156$, $\gamma = 0.8$, $\alpha = 0.1$, $n_T = 10$, $\varsigma = 0.0001$, $\varepsilon = 0.0001$, $\delta_d = 0.025 \|\vQ\bellman_0\|$, $\beta = 1$, $\overline{v}_\mathrm{taxi} = 22$ miles per hour, $\tau = 0.0001$, and $t_\mathrm{RHC} = 10$. We train a reward model using the data from October 29$\th$, 2016 with a total of $91,165$ customer requests. Then, we collect $54,115$ customer requests from October 30$\th$, 2016, which we reveal real-time to the H-TD\textsuperscript{2} and RHC dispatch algorithms. In total, the H-TD\textsuperscript{2} algorithm has a total customer waiting time of \SI{501} hours an improvement of \SI{26}{\percent} over the RHC baseline of a cumulative customer waiting time of \SI{684} hours. This result demonstrates the robustness of adaptive algorithms to irregular events.

\section{Conclusion}
In this work, we present H-TD\textsuperscript{2}, a novel approach to taxi dispatch that exploits the natural topology of transportation network systems with hybrid behavior: a fleet of taxis compute lightweight policy updates locally and, episodically, the policy of all taxis are reset with a centralized batch update from the dispatch center. We derive a regret bound on the local policy estimation that introduces a trigger condition to the batch update, permitting the user to explicitly specify the computational expense/policy bounded sub-optimality trade-off. Unlike state-of-the-art taxi dispatch methods, this algorithm is adaptive, model-free and coordinated, permitting the fleet to adapt in a flexible manner to regular and non-regular events. Compared to a receding horizon control baseline, the proposed algorithm decreases average customer waiting time by \SI{50}{\percent} on a synthetic dataset across a wide range of system parameters, and by \SI{26}{\percent} on real customer data from the city of Chicago taxi dataset over irregular customer distributions during Game 5 of the 2016 Baseball World Series. In future work, we propose (i) testing H-TD\textsuperscript{2} in higher-fidelity urban environments by implementing a data-driven road model and (ii) coupling H-TD\textsuperscript{2} with a low-level motion planner, e.g.~\cite{Riviere_2020}. 

\section*{Acknowledgment}
The authors thank the feedback from colleagues in the Data-driven Intelligent Transportation workshop (DIT 2018, held in conjunction with IEEE ICDM). The authors also thank the reviewers for their helpful feedback and comments. The use of binary log-linear learning was proposed by Salar Rahili in an early version of our work. The work is funded in part by the Raytheon Company and the Jet Propulsion Laboratory. 

\bibliographystyle{IEEEtran}
\bibliography{IEEEabrv,papers}

\vspace{-1cm}
\begin{IEEEbiographynophoto}{Benjamin Riviere} is a PhD student at California Institute of Technology. He received the B.S. in Mechanical Engineering from Stanford University in 2017 and the M.S. in Aeronautics from Caltech in 2018. His research interests are machine learning and network control theory with applications in robotic and transportation systems. 
\end{IEEEbiographynophoto}

\begin{IEEEbiographynophoto}{Soon-Jo Chung} received the B.S. degree (summa cum laude) from Korea Advanced Institute of Science and Technology, Daejeon, South Korea, in 1998; the S.M. degree in aeronautics and astronautics; and the Sc.D. degree in estimation and control from Massachusetts Institute of Technology, Cambridge, MA, USA, in 2002 and 2007, respectively. He is the Bren Professor of Aerospace and Control and Dynamical Systems, and Jet Propulsion Laboratory Research Scientist in the California Institute of Technology. Dr. Chung was on the faculty of the University of Illinois at Urbana-Champaign (UIUC) during 2009-2016. His research focuses on spacecraft and aerial swarms and autonomous aerospace systems, and in particular, on the theory and application of complex nonlinear dynamics, control, estimation, guidance, and navigation of autonomous space and air vehicles. Dr. Chung is received the UIUC Engineering Deans Award for Excellence in Research, the Beckman Faculty Fellowship of the UIUC Center for Advanced Study, the U.S. Air Force Office of Scientific Research Young Investigator Award, the National Science Foundation Faculty Early Career Development Award, and three Best Conference Paper Awards from the IEEE, and the American Institute of Aeronautics and Astronautics. He is an Associate Editor of IEEE Transactions on Robotics, IEEE Transactions on Automatic Control, and Journal of Guidance, Control, and Dynamics.
\end{IEEEbiographynophoto}

\appendices

\section{Fleet Simulator}
\label{appendix:fleet_simulator}
In this section we describe the implementation of the fleet simulator described in Algorithm~\ref{algo:fleet_control}. In particular, the fleet is initialized with uniform random initial positions in the state space and in the dispatch operating mode (Line 1). The customer requests are either synthetically generated in Gridworld or drawn from the Chicago city taxi dataset (Line 3). The taxis are assigned to customers with a distributed matching algorithm (Line 4). After the taxi's choose a dispatch action, the environment updates as follows:

The dynamical model of the $i\th$-taxi servicing customer $\vc^k$ is defined as: 
\begin{alignat}{2}
    \vp^i_t &= 
    \begin{cases}
        \vp^i_{t^k_r} + \frac{t - t^k_r}{t^{k,i}_p- t^k_r} (\vp^{k,p} - \vp^i_{t^i_r}) & t^k_r \leq t < t^{k,i}_p\\ 
        \vp^{k,p} + \frac{t - t^k_p}{t^k_d} (\vp^{k,d} - \vp^{k,p}) & t^{k,i}_p \leq t < t^{k,i}_p+ t^k_d \\ 
    \end{cases} 
\end{alignat}
where this model is evaluated for all servicing taxis, i.e. $\forall i\in\sI_s$, $t^{k,i}_p= t^k_r + \feta(\vp^{k,p},\vp^i_{t^k_r})$ is the pickup time of customer $k$ by taxi $i$, and $\feta$ is the estimated time of arrival function for the environment. 

The dynamic model of the $i\th$-taxi dispatched by dispatch action $a^i_t$ is given by: 
\begin{alignat}{2}
    \vp^i_{t+1} &= \vp^i_t + \frac{\Delta_t}{\feta(\vu^i_t,\vp^i_t)} (\vu^i_t - \vp^i_t) 
\end{alignat}
where this model is evaluated for all free agents, i.e. $\forall i\in\sI_f$, $\Delta_t$ is the simulation timestep, and $\vu^i_t$ is the desired position vector received from the dispatch algorithm.

The \textit{Estimated Time of Arrival (ETA)}, $\feta$ accepts arbitrary position vectors, returns a scalar time value and is defined as:
\begin{align}
    &\feta(\vp^1,\vp^2) = \frac{ \| \vp^1 - \vp^2 \| }{ \overline{v}_\mathrm{taxi} } 
\end{align}
where $\overline{v}_\mathrm{taxi}$ is the average taxi speed, computed offline as an input parameter. 

\begin{remark}
Our proposed dispatch algorithm is agnostic to choice of models $\ff_f$, $\ff_s$, $\feta$ so these can be augmented with a road network models or a data-driven approach. 
\end{remark}

\end{document}